\documentclass[a4paper, 10pt]{article}
\usepackage{mathtools}
\usepackage{amssymb}
\usepackage{algorithmic,algorithm2e,float}
\usepackage{multicol}
\usepackage{amsfonts}
\usepackage{graphics}
\usepackage{enumitem}
\newcommand{\R}{{\mathbb R}}
\newcommand{\PP}{{\mathcal P}}

\newtheorem{theorem}{Theorem}
\newtheorem{lemma}[theorem]{Lemma}
\newtheorem{remark}[theorem]{Remark}
\setlist[itemize]{leftmargin=*}
\begin{document}
\begin{center}

\textbf{Chaos control of Hastings-Powell model by combining chaotic motions}
\vspace{5mm}

\textbf {Marius-F. Danca}\\
{Romanian Institute of Science and Technology,\\
400487 Cluj-Napoca, ROMANIA\\
danca@rist.ro}

\textbf {Joydev Chattopadhyay}\\
{Agricultural and Ecological Research Unit Indian Statistical Institute\\
203, B. T. Road, Kolkata - 700 108, INDIA\\
joydev@isical.ac.in}


\end{center}

\begin{abstract}
In this paper, we propose a Parameter Switching (PS) algorithm as a new chaos control method for the Hastings-Powell (HP) system. The PS algorithm is a convergent scheme that switches the control parameter within a set of values while the controlled system is numerically integrated. The attractor obtained with the PS algorithm matches the attractor obtained by integrating the system with the parameter replaced by the averaged value of the switched parameter values. The switching rule can be applied periodically or randomly over a set of given values. In this way, every stable cycle of the HP system can be approximated if its underlying parameter value equalizes the average value of the switching values. Moreover, the PS algorithm can be viewed as a generalization of Parrondo's game, which is applied for the first time to the HP system, by showing that losing strategy can win: ``losing + losing = winning''. If ``loosing'' is replaced with ``chaos'' and, ``winning'' with ``order'' (as the opposite to ``chaos''), then by switching the parameter value in the HP system within two values, which generate chaotic motions, the PS algorithm can approximate a stable cycle so that symbolically one can write ``chaos + chaos = regular''. Also, by considering a different parameter control, new complex dynamics of the HP model are revealed.


\end{abstract}

\textbf{In Refs. \cite{dan} and \cite{dan2} it is shown that the attractors of a chaotic system, depending on a real parameter, can be numerically approximated by switching the control parameter with the PS algorithm in some deterministic or random manner, while the underlying initial value problem is numerically integrated. In this way, the obtained attractor approximates the attractor obtained via the parameter control using the average of the switched values. If the switching parameter values correspond to chaotic behaviors and the attractor generated by the PS algorithm is a stable cycle, one obtains a chaos control-like (similarly one can obtain a chaos anticontrol-like) schemes. In Ref. \cite{par1}, it is shown that alternating randomly or deterministically the loosing gains of two games, one can actually obtain a winning game with a positive gain; in other words, ''two ugly parents can have a beautiful child'' (Zeilberger, when receiving the 1998 Leroy P. Steele Prize). If one switches, for example, the control parameter within a set of two values which generate chaotic behaviors, denoted with $chaos_1$ and $chaos_2$, by choosing a suitable switching rule, one can obtain a stable cycle denoted as $order$. Symbolically, this can be written as $chaos_1+chaos_2=order$, i.e. Parrondo's paradox variant to controlling chaos. In this paper, we show that the PS algorithm can be utilized as a chaos control-like scheme for the Hastings-Powell model. Also, we show that the PS algorithm can be considered as a generalization of the Parrondo paradox utilized for chaos control in this same model. }

\section{INTRODUCTION}\label{intro}
The evidence of chaos in real systems has led to the need for control of chaos, usually by replacing  chaotic dynamics through stability and stable cycles, which is becoming a fascinating subject of research in different fields including especially ecological systems. To date, many researchers have proposed how to control chaos in e.g. food-chain models by incorporating several biological means. In fact, some biological phenomena like imposition of a population floor \cite{rux1,rux2}, addition of refugia \cite{eis}, omnivory \cite{hast2}, intraspecific density dependence \cite{xu}, toxic inhibition \cite{chat1}, spatial effect \cite{mai}, cascading migration \cite{chat4}, predator feeding switching \cite{pal} can facilitate the control of chaos.

Since the pioneering works of \cite{lot1,lot2} and Volterra \cite{volt}, that model the oscillations of two-specie predator-prey populations and present a mechanisms for neutrally stable limit cycles, the field of mathematical ecology has flourished significantly, which studies ecological models of interacting populations of different species.

Predator–-prey population cycles represent a great deal of fascination for the scientific world of animal naturists, becoming a major research subject in ecology since the last century, with the first important scientific reports being \cite{may,sel,el} or \cite{sm} (see more references in \cite{vass}).

Despite the fact that in some early works the investigators considered that the systems, or the system parameters, inducing chaos are unrealistic from biological or systems point of view, Hastings and Powell \cite{hast} introduced a thritrophic chaotic model (HP model), which is an extension of the two-variable Rosenzweig-MacArthur model \cite{ros,ros2}. The system describes the food-chain interaction within an ecosystem of three species with Type II functional responses. It is an intriguing model which can be viewed as an approximation for large classes of biological systems. They showed that the system possesses strange attractors. Since their nonlinear multi-trophic model was introduced, theoretical ecologists started to analyze the subtle dynamics of these kind of systems (see e.g. \cite{kleb,rin,mc,kus} or \cite{str}, which is an accessible mathematical introduction of complex dynamics with many biological examples).

The HP system is modeled by the following system
\begin{equation}\label{mod0}
\begin{array}
[c]{ll}
\frac{dX}{dT}= R_0X(1-X/K_0)-C_1F_1(X)Y,\\[10pt]
\frac{dY}{dT}=F_1(X)Y-F_2(Y)Z-D_1Y,\\[10pt]
\frac{dZ}{dT}=C_2F_2(Y)Z-D_2Z,
\end{array}
\end{equation}

\noindent where
\begin{gather}
F_i(U)=A_iU/(B_i+U),~~~ i=1,2,
\end{gather}

\noindent represent a Holling type II functional response of both consumer species \cite{hol} with $B_i$, which is the half-saturation constant satisfying $F_i(U)|_{U=B_i}=A_i/2$; $T$ represents time, $R_0$ is the intrinsic growth rate, $K_0$ is the carrying capacity of species $X$, constants $C_1^{-1}$ and $C_1$ are conversion
rates of the prey to predator of species $Y$ and $Z$ respectively, $D_1$ and $D_2$ are constant death rates for species $Y$ and $Z$ respectively, constants $A_{1,2}$ and $B_{1,2}$ parameterize the saturation functional response and $B_i$ is the prey population level where the predation rate per prey unit is half of its maximum value. Due to the set of $10$ parameters, which imply difficult analysis, the system is transformed into a nondimensional form by the linear transformation (given in \cite{hast})
\begin{gather}
x_1=\frac{X}{K_0},~~~x_2=\frac{C_1Y}{K_0},~~~x_3=\frac{C_1Z}{C_2K0},~~~t=R_0T,
\end{gather}

\noindent yielding the following system with only $6$ parameters
\begin{equation}\label{mod}
\begin{array}
[c]{cl}
\dot{x}_{1}= & x_1(1-x_1)-g_1(x_1)x_2,\\
\dot{x}_{2}= & g_1(x_1)x_2-g_2(x_2)x_3-d_1x_2,\\
\dot{x}_{3}= & g_2(x_2)x_3-d_2x_3,
\end{array}
\end{equation}

\noindent where $g:\R\rightarrow\R$ is a real function of the form

$$
g_i(x)=\frac{a_ix}{1+b_ix}.
$$

\noindent The original parameters are: $a_1=5, b_1=5, a_2=0.5, a_2=0.1, b_2=2, d_2=0.01$. In the previous papers, the bifurcation parameter is $b_1$. In this paper, we consider $p:=d_1$ as the control parameter. It should be mentioned that similar results to those presented in this work are obtained if $d_2$ is used as the control parameter.
We show that the chaotic behavior appears in regions of the parameter space different to e.g \cite{hast,yod1,yod2} or \cite{ston}. Also, the bifurcation diagram (Fig. \ref{fig2} a) reveals  much richer dynamics than the case when $b_1$ is used as the bifurcation parameter. A typical stable periodic motion, whose reverse ``tea-cup'' shape characterizes the great majority of HP's chaotic attractors, is drawn with tubes in Fig. \ref{fig0}. The colors mark the speed on the trajectory, from the highest (red) to the lowest (yellow).

Since in realistic cases it is almost impossible to distinguish quasiperiodic from chaotic behavior using only classical methods of period analysis, one usually use tools like the correlation dimension or Lyapunov coefficients \cite{vot}, which is beyond the scope of this paper. Instead, one can consider that, beside chaotic motion revealed in previous works, the HP system presents periodic, and possibly quasiperiodic oscillations, often of mixed-mode type, with periodic or nearly periodic oscillations (in \cite{xia}, the authors revealed some quasiperiodic motion, via Poincar\'{e} section). As known, mixed-mode oscillations, found first in chemical reactions (one of the most known example being the classic Belousov-Zhabotinsky reaction \cite{masel}), are oscillations (periodic or not) in which there is an alternation between oscillations with clearly separated large and small amplitudes (see for example the comprehensive survey \cite{mmo}). These oscillations were noticed for over thirty years existing many dynamic systems. We found possible mixed-mode oscillations in the HP system with $q=0.36$ (see the phase plot and time series in Fig.\ref{fig2} a). These mixed-mode oscillations, as indicated by colors in Fig. \ref{fig0}, are typical to ODEs with slow-fast dynamics.

In the current literature, there are only a few results on chaos control for this kind of systems (see e.g. \cite{gom}, where the chaos control is achieved by stabilizing two saddle orbits via the OGY method \cite{og}). Recently, we proposed the PS algorithm \cite{dan} and explored the possibility of obtaining desired attractors by switching a key control parameter. Using this scheme, we can obtain a stable cycle of the HP system if the control parameter is switched within a set of values which generate chaotic motions. Compared to the known chaos control techniques (such as the OGY method), which force some unstable periodic orbits to become stable, the PS algorithm allows to approximate any real stable cycle (also chaotic attractors) by simple switching of the control parameter within a chosen set, without influencing the intrinsic complexity of the underlying system dynamics. Also, contrarily to the known control and anticontrol algorithms, which modify some unstable periodic orbits in order to become stable, the PS algorithm allows to obtain one of the possible motions of the underlying system, without modifying its properties.

Because the PS algorithm can be applied in a deterministic (periodic) manner \cite{dan} and also in some random manners \cite{dan2}, we show in this paper for the first time that the PS algorithm can be easily implemented for chaos control in the HP model, and also that it can be used to explain what could happen in some natural systems (such as the tritrophic food chain HP system) when periodical or random switchings between chaotic motions are applied. We show that it can lead to some stable cycles or, reversely, can explain why switching between stable periodic evolutions can drive the underlying system to evolve chaotically.

\section{PARAMETER SWITCHING ALGORITHM}\label{PS}

Consider a class of systems modeled by the following Initial Value problem (IVP)
\begin{equation}\label{e0}
\dot x(t)=f(x(t))+pAx(t),\quad t\in I=[0,T], \quad x(0)=x_0,
\end{equation}
\noindent where $T>0$, $x_0\in \mathbb{R}^n$, $p\in \mathbb{R}$ is the control parameter, $A\in L(\mathbb{R}^n)$, and $f : \mathbb{R}^n \rightarrow\mathbb{R}^n$ a nonlinear function. The IVP (\ref{e0}) models a majority of continuous-time nonlinear and autonomous dynamical systems depending on a single real control parameter $p$, such as the Lorenz system, R\"{o}sler system, Chen system, Lotka–-Volterra system, Rabinovich–-Fabrikant system, Hindmarsh-Rose system, L\"{u} system, some classes of minimal networks, and many others. For the HP system (\ref{mod}), $n=3$, and can one choose $p=d_1$, with

\[
f(x)=\left(
\begin{array}{c}
x_{1}(1-x_{1})-g_1(x_1)x_2 \\
g_1(x_1)x_2-g_2(x_2)x_3\\
g_2(x_2)x_3-c_2x_3%
\end{array}%
\right) ,~~A=\left(
\begin{array}{ccc}
0 & 0 & 0 \\
0 & -1 & 0 \\
0 & 0 & 0%
\end{array}%
\right) .
\]

The PS algorithm applied to (\ref{e0}) allows to approximate any desired solution  \cite{dan}. Thus, by choosing a finite set of $N>1$ parameters values, $\mathcal{P}_N=\{p_1,p_2,...,p_N\}$, the PS algorithm switches the parameter $p$ within $\PP_N$ for a relatively short time subintervals $I_{i,j}$, $i=1,2,...,N$, $j=1,2,...$, such that $I=\bigcup_{j}\bigcup_{i=1}^NI_{i,j}$, while the underlying IVP is numerically integrated (Fig. \ref{fig1}). The resulted ``switched'' solution approximates the ``averaged'' solution, which is obtained when the parameter $p$ is replaced with the average of the switched values, as follows

\begin{equation}\label{p}
p^*:=\frac{\sum_{i=1}^Nm_ip_i}{\sum_{i=1}^Nm_i},
\end{equation}

\noindent where $m_i$, $i=1,2,...,N$, are some positive integers, and $p_i$ are weights. The ``switching'' equation (related to the PS algorithm) has the following form
\begin{equation}\label{e1}
\dot x(t)=f(x(t))+p(t)Ax(t),\quad t\in I=[0,T],\quad x(0)=x_0,
\end{equation}

\noindent where $p:I\rightarrow\PP_N$ is a piece-wise constant function that switches its values in the subintervals $I_{i,j}$,
$p(t)=p_i,~~~ t\in I_{i,j},~~~i\in\{1,2,...,N\},~~~j=1,2,...$ (Fig. \ref{fig1}).

\noindent For simplicity, hereafter, the index $j$ will be dropped.

\noindent The ``averaged'' equation of (\ref{e0}), obtained for $p$ being replaced with $p^*$ given by (\ref{p}), is
\begin{equation}\label{e2}
\dot {\bar x}(t)=f(\bar x(t))+p^*A\bar x(t),\quad t\in I=[0,T],\quad \bar x(0)=\bar x_0.
\end{equation}

Throughout, the following assumptions are made.

\noindent \textbf{Assumption H1}.
$f$ satisfies the Lipschitz condition.

\noindent\textbf{Assumption H2}. The initial conditions $x_0$ and $\overline{x}_0$ of (\ref{e1}) and (\ref{e2}) respectively, belong to the same basin of attraction $\mathcal{V}$ of the solution of (\ref{e2}).

Then, the global error (difference between the solutions of (\ref{e1}) and (\ref{e2})) is given by the following lemma \cite{dan,mao}.

\begin{lemma}\label{lem}
Consider the IVPs (\ref{e1}) and (\ref{e2}). Given any close initial conditions $x_0, \overline{x}_0\in \mathcal{V}$, the switched solution approximates the averaged solution.

\end{lemma}

The proof is presented in \cite{mao}, where the average theory (see e.g. \cite{ver}) has been utilized, while in \cite{dan} the proof is made via the the global error of Runge-Kutta method.

Next, consider the following reasonable assumption regarding the notion of attractor utilized, and necessary to numerically implement the PS algorithm.

\noindent \textbf{Assumption H3}. To every $p$ value, for a given initial condition $x_0$, there corresponds a unique solution and, therefore, a single attractor, denoted by $A_p$, which can be approximated numerically by its $\omega$-limit set \cite{foi}, after neglecting a sufficiently long period of transients.

\noindent The following theorem represents the main result concerning the PS algorithm applied to systems modeled by the IVP (\ref{e0})

\begin{theorem}\label{th}
 Every attractor of the system (\ref{e0}) can be numerically approximated by the PS algorithm.
\end{theorem}

\noindent Denote by $A^*$ the ``synthesized attractor'', obtained with the PS algorithm, and by $A_{p^*}$ the ``averaged attractor'', obtained for $p$ being replaced with $p^*$ given by (\ref{p}).

To obtain a desired attractor corresponding to some value $p$, we replace $p^*$ with $p$ in (\ref{p}) and choose a set $\PP_N$ with the weights $m_i$, $i=1,2,...,N$, such that (\ref{p}) is verified. Then, by applying the PS algorithm, the obtained switched attractor $A^*$ will approximate the searched (averaged) attractor $A_{p}$.

Theorem \ref{th} means that by choosing some value $p$, there exists the attractor $A_{p}$ (see Assumption H3) and a set of $N>1$ parameters $\PP_N$, such that $p^*=p\in (p_{min},p_{max})$ with the weights $m_i$, $i=1,2,...,N$, and $p^*$ given by the relation (\ref{p}). Then, as stated by Theorem \ref{th}, $A_{p^*}$, will be approximated by the attractor $A^*$, generated by the PS algorithm.

To numerically implement the PS algorithm, we choose a numerical method with fixed step-size $h$ to integrate the IVP (\ref{e0}), a set of switching parameters $\PP_N$ with weights $m_i$, $i=1,2,..., N$, and a uniform partition of the time interval $I$ in the adjacent subintervals $I_i$, $i=1,2,...,N$, of lengths $m_ih$. With these ingredients, for a fixed step size $h$, the PS algorithm can be expressed as
\begin{equation}\label{scheme}
[m_1p_1,m_2p_2,...,m_Np_N],
\end{equation}

\noindent which means that in the first time subinterval $I_1$, for the first $m_1$ steps of length $h$, the integration is made with $p=p_1$. Then, in the next subinterval $I_2$, for $m_2$ steps, the integration is made with $p=p_2$, and so on, until the last subinterval $I_N$, of length $m_Nh$, where the integration is made with $p=p_N$. Next, the algorithm repeats on another set of $N$ subintervals $I_i$, $i=1,2,...,N$, and so on, until the interval $I$ is entirely covered (see Algorithm 1).

\vspace{3mm}
\begin{remark} \label{remus}
\vspace{-\baselineskip}
\begin{itemize}
\item[(i)]
Taking into account of the convexity of the relation (\ref{p}) (if one denotes $\alpha_i={m_i}/{\sum_{k=1}^Nm_k}$, then $\sum_{i=1}^N\alpha_i=1$, and $p^*=\sum_{i=1}^N\alpha_ip_i$), the only necessary condition to approximate some attractor $A_p$ is to choose $\PP_N$ such that $p\in(p_{min},p_{max})$, with $p_{min}=min\{\PP_N\}$ and $p_{max}=max\{\PP_N\}$. The order of $p_i$ in (\ref{scheme}) is irrelevant, with $p_{min}=p_1$ and $p_{max}=p_N$.
\item[(ii)]
The above-mentioned convexity implies a robustness-like property of the PS algorithm: for every set $\PP_N$, $A^*$ will be situated ``between'' the attractors $A_{p_{min}}$ and $A_{p_{max}}$, with the order being induced by the natural order of the real parameter $p$ (see the sketch in Fig. \ref{fif_fig}).
\item[(iii)] However, the order of parameter values $p_i$ in the scheme (\ref{scheme}) is not important.
\end{itemize}
\end{remark}

The size of $h$ is required only by the convergence of the underlying numerical method utilized in the PS algorithm, for example the standard Runge-Kutta scheme utilized in this paper. If one considers the simplest case of the scheme $[m_1p_1,m_2p_2]$ with $p_{1,2}$ belonging to different chaotic windows in the parameter space (visualized in the bifurcation diagram), there exists at least one periodic stable window between these chaotic windows. Then, by a suitable choice of weights $m_{1,2}$, $p^*$ can be localized inside one of these periodic windows (Remark \ref{remus} ii) and, therefore, the PS acts as a control-like algorithm. Conversely, if $p_{1,2}$ are situated in different periodic windows, the weights $m_{1,2}$ can be chosen such that $p^*$ belongs to a chaotic window and the PS algorithm can be considered as an anticontrol-like algorithm. This also happens in the general case with $N>2$ parameters.

\begin{remark}\label{re}
\noindent
As is known, attractors present continuous dependence on a parameter which, roughly
speaking, means that the dependence of the solution of the IVP (\ref{e0}) on the parameter $p$
is continuous as long as the function $f$ is continuous (see e.g. \cite{hal} or \cite{perk}, p. 83).
Therefore, contrarily to reasonable expectations, with relatively small steps size $h$, the PS algorithm does not affect the solution continuity.
\end{remark}

As seen bellow, the scheme ($\ref{scheme}$) can be applied periodically, with the period $(m_1+m_2+...+m_N)h$. For example, for a given $h$, the scheme $[2p_1,3p_2]$ means that while the IVP is integrated with the PS algorithm, one obtains the following $5h$-periodic parameter sequence: $(p_1,p_1,p_2,p_2,p_2),(p_1,p_1,p_2,p_2,p_2),...$. Moreover, the scheme (\ref{scheme}) can also be applied randomly. For example, for a random uniformly distributed sequence of the time subintervals $I_i$, $i=1,2,...,N$, and $p_i$ respectively (see Algorithm 2), the averaged value, denoted now with $\overline{p}^*$, is determined by
\begin{equation}\label{pp}
\overline{p}^*:=\frac{\sum_{i=1}^Nm'_ip_i}{\sum_{i=1}^Nm'_i},
\end{equation}
\noindent where, $m'_{i}$ is the total number of switchings of $p_{i}$ when the integration ends. On a sufficiently large integration interval $I$, due to the uniformly distributed choice of subintervals $I_i$, $i=1,2,...,N$, $m'_i=n_im_i$, and after a sufficient large integration steps number $n_i$ verify the relations: $n_1\approx n_2\approx...\approx n_N:=n$ with a small error. Next, due to the convexity (Remark \ref{remus} (iii)), $\overline{p}^*=\frac{\sum _{i=1}^N m'_ip_i}{\sum_{i=1} m'_i}=\frac{\sum _{i=1}^N n_im_ip_i}{\sum_{i=1}^N n_im_i}\approx \frac{n\sum_{i=1}^N m_ip_i}{n\sum_{i=1}^N m_i}=p^*$. Other random ways to implement the PS algorithm can be imagined.

The numerical limitations of the PS algorithm reside mainly in the computational errors (truncation error, rounding error, size of $h$, $p$ rational nonterminating number, etc.). For example, the detailed $D$ in Figs. \ref{fig3} and \ref{fig6}, reveal some relatively larger differences between the two attractors. These may be due to the utilized numerical scheme, but also due to the system dynamics. As can be seen in Fig. \ref{fig0}, there are some peaks where the system speed along the trajectory is a few times lower than that on the straight portions. Also, larger values of $m_i$ can lead to some differences between the averaged attractor and the synthesized attractor. For example, if the values of $m_i$ are too large, the averaged attractor presents, within the underlying short periods of time $I_{i,j}$, the tendency to converge towards the attractor for the corresponding $p_i$ \cite{danx}.

\section{FINDING STABLE CYCLES OF THE HP SYSTEM}

\begin{figure}
\begin{center}
  \includegraphics[clip,width=1\textwidth] {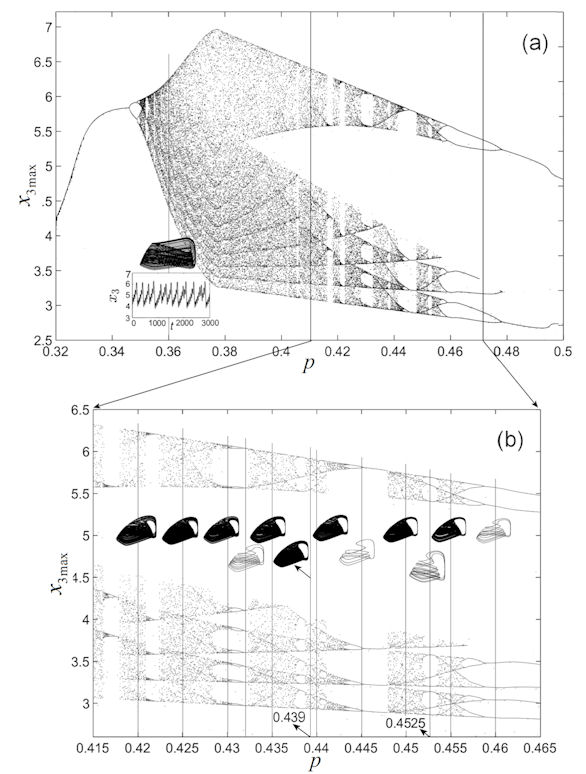}
\caption{Bifurcation diagram of the HP system (\ref{mod}); a) The detail reveals a mixed motion corresponding to $p=0.36$; b) Distribution of the attractors.}
\label{fig2}
\end{center}
\end{figure}

\begin{figure}
\begin{center}
  \includegraphics[clip,width=0.45 \textwidth] {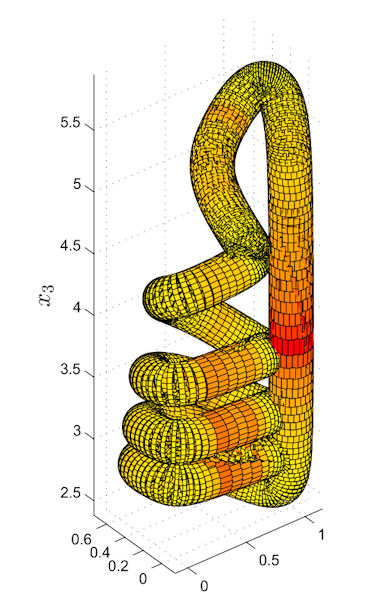}
\caption{Tubular stable cycle of the HP system. Red color indicates highest speed and yellow color the lowest speed along the attractor.}
\label{fig0}
\end{center}
\end{figure}

\begin{figure}
\begin{center}
  \includegraphics[clip,width=0.8\textwidth] {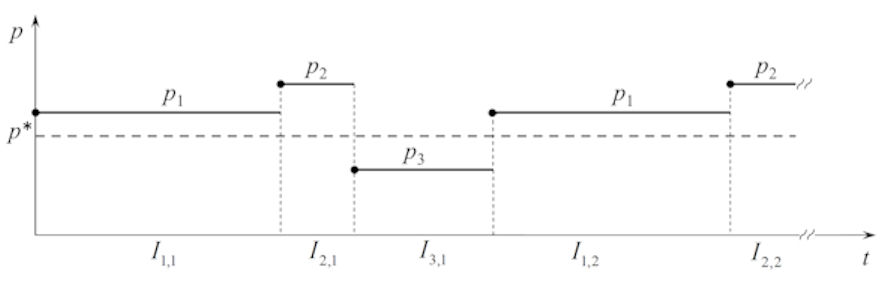}
\caption{Sketch of the $p$ switching for the case of $N=3$.}
\label{fig1}
\end{center}
\end{figure}

\begin{figure}
\begin{center}
  \includegraphics[clip,width=0.55\textwidth] {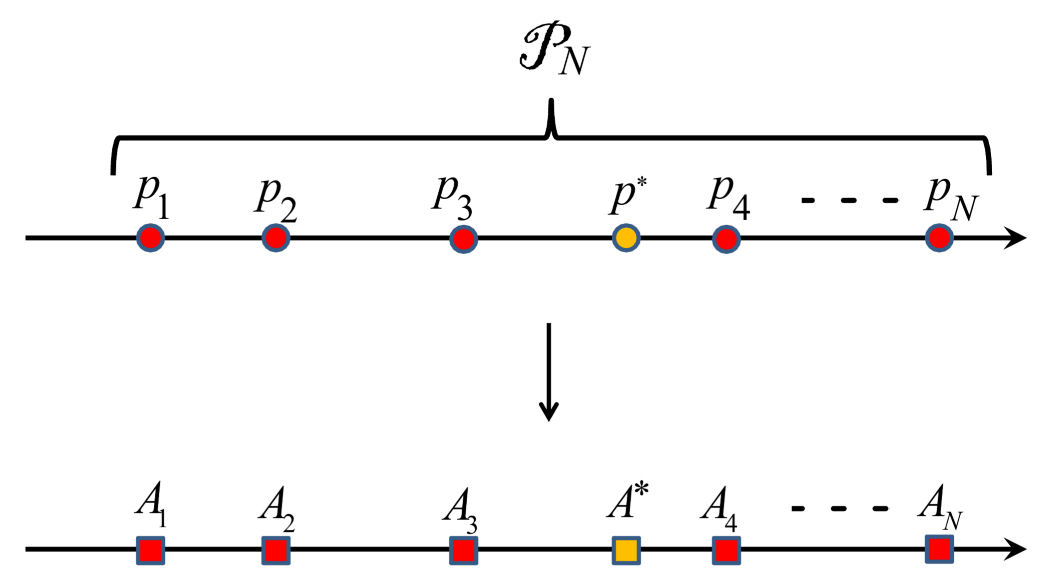}
\caption{The convexity of the PS algorithm (sketch).}
\label{fif_fig}
\end{center}
\end{figure}

In this section, we apply the PS algorithm to approximate some representative stable cycles of the HP system (\ref{mod}). To facilitate the choice of the parameters, the bifurcation diagram is utilized (see Fig.\ref{fig2} b, where the considered $p$ values are plotted). The numerical method used here to solve the system is the standard Runge-Kutta (written in Matlab language by following Algorithms \ref{al1},\ref{al2}). The randomness is based on the pseudorandomness of the Matlab functions. Because competitions between populations occur on a larger time scale, the integration interval $I$ has been chosen in the order of $1E3$. However, though it is known that, if the solution is unique (due to the Lipschitz condition), then the solution is computable over its lifespan (the maximal interval on which the solution exists), an accurate long-time solution remains to be a challenge for the classical numerical methods (see e.g. \cite{col,hai}). A practical way of considering the validity of the numerical results for a given system, such as the HP system, is to use at least two different methods to solve the same problem. If the two solutions agree, then we can have some confidence about the computed solutions. here, the results obtained with the standard Runge-Kutta method have been confronted successfully with the implemented Matlab routines for ODEs. Transients have been removed. In all simulations, both attractors $A^*$ and $A_{p^*}$ are overplotted to underline the perfect match between the searched attractor $A_{p^*}$ and the approximating attractor $A^*$. The match between $A^*$ (blue plot) and $A_{p^*}$ (red plot) is emphasized by phase overplots, time series or Poincar\'{e} section.

\vspace{3mm}
\noindent \textbf{Periodic scheme (Algorithm 1)}

\begin{figure}
\begin{center}
  \includegraphics[clip,width=0.75\textwidth] {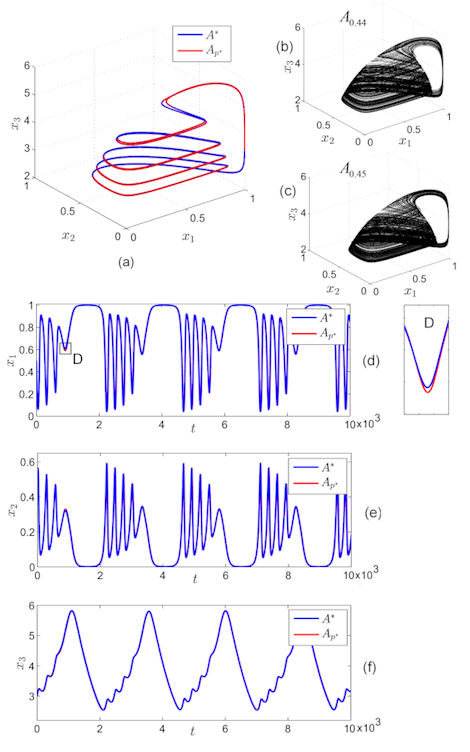}
\caption{Stable cycle in the HP system corresponding to $p=0.445$ obtained with the scheme $[1p_1,1p_2]$, $p_1=0.44$ and $p_2=0.45$; a)Phase overplots of $A^*$ and $A_{p^*}$, for $p^*=0.445$; b) Attractor $A_{0.44}$; c) Attractor $A_{0.45}$; d-f) Overplotted time series of $A^*$ and $A_{p^*}$.}
\label{fig3}
\end{center}
\end{figure}

1) \label{ex1}Suppose one wants to obtain the stable cycle corresponding to $p=0.445$ (close to a reverse period-doubling point) with $N=2$ and the alternation between $p_1$ and $p_2$ given by the scheme $[1p_1,1p_2]$ with $p_1= 0.44$ and $p_2=0.45$. $p_1$ and $p_2$ are chosen such that $p\in(p_1,p_2)$ (see Remark \ref{remus} i) satisfying (\ref{p}). In this case, the average attractor to be approximated is $A_{p^*}$ with $p^*={(1\times 0.44+1\times 0.45)}/{(1+1)}=0.445$, which will be approximated by $A^*$ obtained with the PS algorithm (Fig.\ref{fig3} a). The match between the two attractors is revealed by the time series (Fig.\ref{fig3} d-f). Even though the attractors $A_{0.44}$ and $A_{0.45}$ are chaotic (Fig.\ref{fig3} b,c), the obtained attractor, $A_{0.445}$, represents a stable cycle.

\begin{figure}
\begin{center}
  \includegraphics[clip,width=0.95\textwidth] {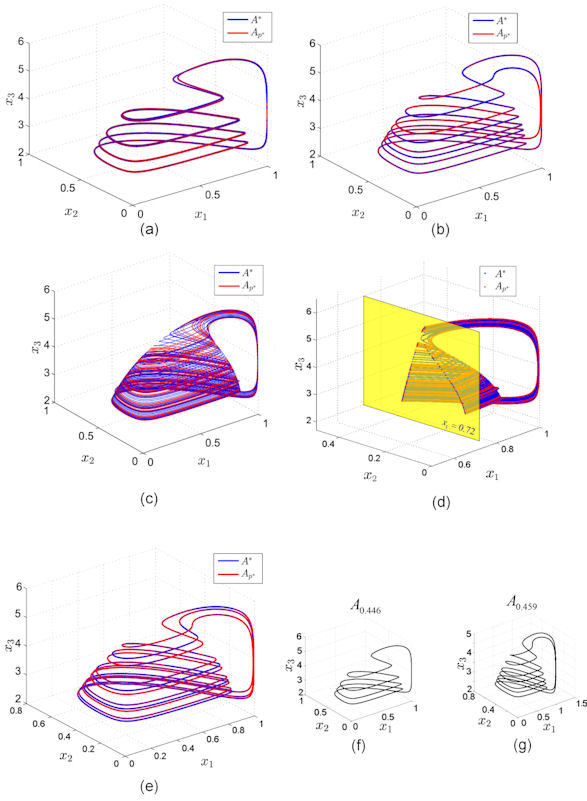}
\caption{a) Stable cycle corresponding to $p=0.445$, obtained with the scheme $[1p_1,1p_2,1p_3,3p_4,4p_5]$, and $\PP_5=\{0.42, 0.425, 0.435, 0.45, 0.455\}$; Phase overplots of $A^*$ and $A_{p^*}$; b) Stable cycle with a higher period corresponding to $p=0.432$, obtained with the scheme $[1p_1,1p_2]$, $p_1=0.425$ and $p_2=0.432$; c) Chaotic attractor corresponding to $p=0.45$ obtained with the scheme $[2p_1,1p_2]$, $p_1=0.445$ and $p_2=0.46$; Phase overplots; d) Poincare section with the plane $x_1=0.72$ of the overplotted chaotic attractors in Fig. c; e) Stable cycle corresponding to $p=0.4525$ obtained with the scheme $[1p_1,1p_2]$, $p_1=0.446$ and $p_2=0.459$ (phase overplots); f) The stable cycle $A_{0.446}$; g) The stable cycle $A_{0.459}$.}
\label{fig4}
\end{center}
\end{figure}

2) \label{ex2}The same stable motion can be obtained, for example, using a set $\PP$ with $N=5$. If one chooses $\PP_5=\{0.42, 0.425, 0.435, 0.45, 0.455\}$, a possible set of weights necessary to give $p=0.445$ are $m_1=m_2=m_3=1$, $m_4=3$ and $m_5=4$, then by the scheme $[1p_1,1p_2,1p_3,3p_4,4p_5]$ the PS algorithm will approximate the same stable cycle $A_{0.445}$ (Fig.\ref{fig4} a).

3) \label{ex3}To obtain stable periodic motions with higher periods, for example, to the periodic window around $p=0.43$ (Fig. \ref{fig2} b), it is possible to approximate the stable cycle corresponding to $p=0.432$. A choice is the scheme $[1p_1,1p_2]$ with $p_1=0.425$ and $p_2=0.439$, values which yield $p=p^*=0.432$.

\noindent The synthesized and averaged attractors match very well (Fig.\ref{fig4} b).

4) \label{ex4}Chaotic attractors can also be obtained. For example, to approximate the chaotic attractor corresponding to $p=0.45$, by switching $N=2$ values, one can use the scheme $[2p_1,1p_2]$ with $p_1=0.445$ and $p_2=0.46$ (see Fig. \ref{fig2}b). Due to the asymptotical characteristic of chaotic attractors, the approximation gives weaker results. However, the phase plots (Fig. \ref{fig4} c) and the Poincar\'{e} section with $x_1=0.72$ show a good match, $A^*$ and $A{p^*}$, evolving basically on the same shape (see Fig. \ref{fig4} d, where the attractors $A^*$ and $A_{p^*}$ are plotted by dotted curves to reveal the intersection points).

5) \label{ex5}To obtain some stable cycle, the switchings can be done using stable cycles. For example, to force the system to evolve on the stable limit cycle corresponding to $p=0.4525$ (see Fig. \ref{fig2} b, Fig. \ref{fig4} e), one can apply the scheme $[1p_1,1p_2]$, with $p_1=0.446$ and $p_2=0.459$, which belong to two different periodic windows (Fig. \ref{fig4} f, g).

\vspace{3mm}

\noindent \textbf{Random scheme 2}

\begin{figure}
\begin{center}
  \includegraphics[clip,width=0.75\textwidth] {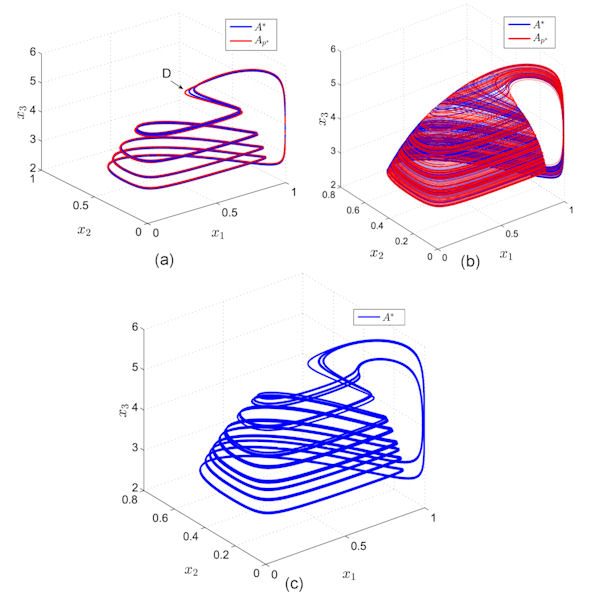}
\caption{Random application of the PS algorithm; a) Stable cycle corresponding to $p=0.445$, obtained with the scheme $[1p_1,1p_2,1p_3,3p_4,4p_5]$, $\PP_5=\{0.42, 0.425, 0.435, 0.45, 0.455\}$; the subintervals $I_{i},~i=1,2,...,5$, with order chosen randomly (phase overplots); b) Chaotic attractor $A^*$ obtained by switching randomly the order of the values of $\PP_5=\{0.42, 0.425, 0.435, 0.45, 0.455\}$ (phase overplots); c) Chaotic attractor obtained by randomly choosing $p$ within the interval $(0.425,0.435)$. }
\label{fig6}
\end{center}
\end{figure}

6) \label{ex6}One can apply randomly the scheme $[1p_1,1p_2,1p_3,3p_4,4p_5]$, with $\PP_5=\{0.42, 0.425, 0.435, 0.45, 0.455\}$ (see Fig. \ref{fig2} b), to obtain the stable cycle corresponding to $p=0.445$, as above. For this purpose, one can choose randomly, for example, in the order of subintervals $I_i$, $i=1,2,...,5$, where the PS algorithm is applied.
 For this purpose, the simplest way is to use a pseudorandom number generator function (for example Matlab's function $randi$). The result (stable cycle) can be seen in Fig. \ref{fig6} a. Now, the relatively small difference between the two attractors in the region $D$ is more accentuated than using the periodic PS algorithm (Fig. \ref{fig3} d).

7) \label{ex7}However, as expected, not even random scheme is chaos control-like algorithm. For example suppose the system evolves under uniformly randomly switching of $p$ within the same set $\PP_5$. After $5E6$ iterations, the occurrence of $p_i$, $i=1,2,...,5$, was: $m_1'=1000568$, $m_2'=999369$, $m_3'=1000343$, $m_4'=1000372$ and $m_5'=999348$. With these values, the relation (\ref{pp}) gives (with eight significant decimals) $\overline{p}^*=0.4370019$, corresponding to a chaotic motion presented in Fig.\ref{fig6} b, but not to a stable cycle as before.

8) \label{ex8}Another possible real situation is that the system suffers random switching of the values of $p$, in some limited interval. Consider $p$ taking random values within an interval including the periodic window around $p=0.43$ (Fig. \ref{fig2} b). For this purpose, consider $p=0.425+rand/100$, which generates pseudorandom numbers $p\in(0.425,0.435)$. After $5E5$ iterations, one obtains a stable cycle with multiple periods (Fig. \ref {fig6} c). Even in this case, due to the difficulty in counting the number of occurrences for each $p$ (weights), it is difficult to find $\overline{p}^*$. However, due to the convexity property (Remark \ref{remus} i), the obtained attractor is one of the real system motions. This example illustrates the robustness-like property of the PS algorithm: for whatever set $\PP_N$ and weights $m_i$, the PS algorithm leads to one of the existing attractors (Remark \ref{remus} ii).

\vspace{3mm}

\emph{Parrondo's game applied to the HP system}

 Let us now consider the counter-intuitive behavior of the game of chance, known as Parrondo's game, or Parrondo's paradox (see e.g. \cite{par1,par2})\footnote{John von Neumann was one of the first mathematicians who proved that there are some kind of games involving bluffing, for which one can have optimal strategies to guarantee the best outcome. Von Neumann's work showed these kind of games to be applicable for example in social behavior, in economics as well in ecology.} in the following form: $losing+losing=winning$. This means that, by alternating two losing strategies in a deterministic way, a winning game can be obtained \cite{par1,par2}. If one replaces $losing$ by $chaos$ and $winning$ by $order$, then the following variant of Parrondo's game is obtained: $chaos_1+chaos_2=order$, i.e. a chaos control-like behavior. Generalizing, one obtains the following form of chaos control-like result, which, written in parrondian words, reads $chaos_1+chaos_2+...+chaos_N=order$ \cite{dan}.
The anticontrol-like algorithm implemented with the PS algorithm can be written as follows: $order_1+order_2+...+order_N=chaos$ (by $chaos_i$ or $order_i$ one understands a chaotic or a stable periodic behavior respectively, corresponding to some value $p_i$). Therefore, the examples (1)-(4) can be characterized in parrondian's terminology as summarized in Table \ref{tab:1}.

\begin{table}
\begin{center}
\scalebox{0.82}{
\begin{tabular}{l c l l c c}
\hline
 Exemple& Parrondo's game& & &PS algorithm & Remarks  \\ [2pt]
\hline
Ex. 1 (Fig. \ref{fig3})&$chaos_1 + chaos_2$&=&$order$& periodic & chaos control-like \\[1pt]
Ex. 2 (Fig. \ref{fig4} a)&$chaos_1 +...+ chaos_5$&=&$order$ &periodic& chaos control-like \\[2pt]
Ex. 3 (Fig. \ref{fig4} b)&$chaos_1 + chaos_2$&=&$order$ & periodic&chaos control-like  \\[1pt]
Ex. 4 (Fig. \ref{fig4} c)&$order_1 + order_2$&=&$chaos$ & periodic&anticontrol-like \\[2pt]
\hline
\end{tabular}
}
\end{center}
\caption{Chaos control-like and anticontrol-like results of the HP system (\ref{mod}) and analogy with Parrondo's game. }
\label{tab:1}
\end{table}

\section{DISCUSSION}

In the present work, we consider the Hastings-Powell model and chose the rate of natural mortality of middle predator as the control parameter. In the real-world situation, it is not always possible to attain some specific values of the rate parameter and the corresponding dynamics of the system. Therefore, suppose one intends to force the system to evolve along the attractor corresponding to some inaccessible value $p^*$, but has access to a set of $N$ parameter values (from experimental/field results) $\mathcal{P}_N$, such as $p^*\in(p_{min},p_{max})$, with $p_{min}=min \{\mathcal{P}_N\}$ and $p_{max}=max \{\mathcal{P}_N\}$. Then, using the PS algorithm one can approximate the attractor corresponding to any intermediary value $p^*$ between $p_{min}$ and $p_{max}$.
It is worthy noting here that if the frequency of the oscillations in prey-middle predator interaction and the frequency of the oscillations in middle predator-top predator interaction are commensurate, then the HP three-species system shows periodic oscillations. Moreover, when such frequencies are incommensurate, chaotic oscillations have been observed. Therefore, the averaged solutions become stable cycles or chaotic cycles depending on the frequencies of the oscillations corresponding to the switched parameter values.
In ecological context, stable dynamics (stable steady states or stable cycles) are desirable as chaotic populations are prone to extinction subjected to stochastic fluctuations. Therefore, the outcomes of the present work are very important and useful even from a management science perspective.

The rate of convergence of PS algorithm could be related to the negative Lyapunov exponents at this point in phase space, which can be calculated from the method in \cite{ek}. In this way, the difference between the actual attractor and the averaged attractor could be related to the $m_i$'s and the negative Lyapunov exponents.
A future task will be to study the convergence of the PS algorithm in the case when $\PP$ contains infinitely many elements.

\vspace{3mm}
\textbf{Acknowledgment}  We thank Calin Gal-Chis for his help.

\begin{figure*}
\begin{center}
  \includegraphics[clip,width=1\textwidth] {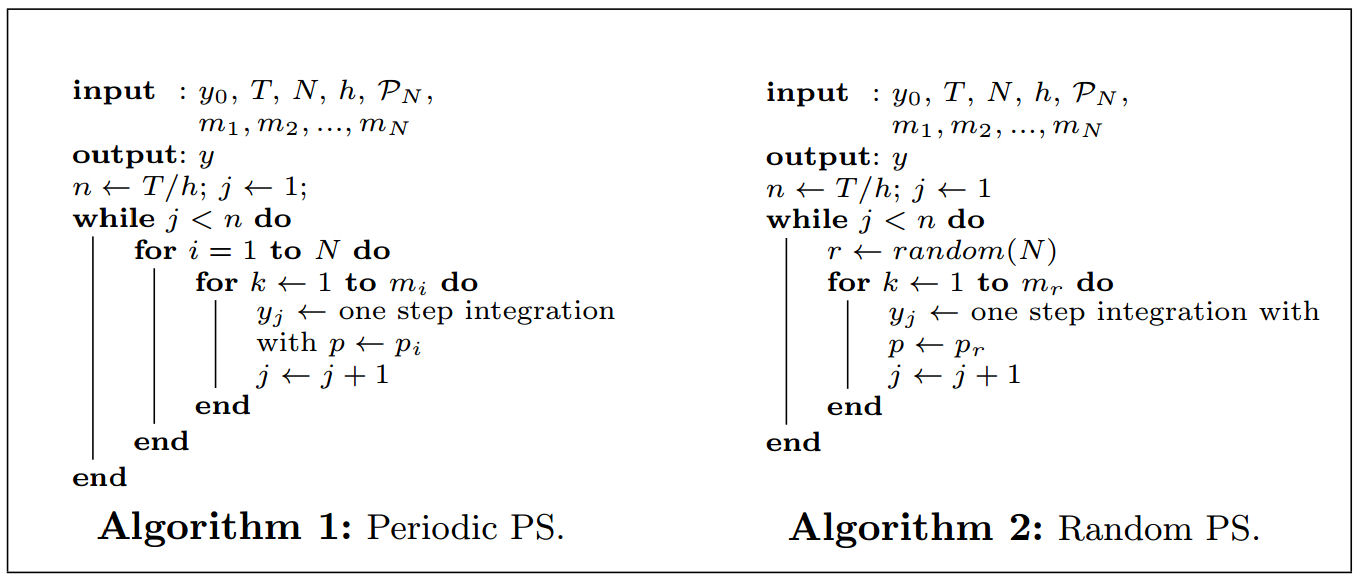}
\caption{}
\end{center}
\end{figure*}


\begin{thebibliography}{100}
\bibitem{chat1} J. Chattopadhayay, and R. Sarkar, Ecol. Model. \textbf{163}, 45 (2003).

	\bibitem{col} P. Collins, and D.S. Grac, J. Univers. Comput. Sci. \textbf{15}, 1162 (2009).

	\bibitem{dan} M.-F. Danca, Commun Nonlinear Sci \textbf{18}, 500 (2013).

\bibitem{dan2}M.-F. Danca, CHAOS, \textbf{18}, 033111 (2008).

\bibitem{danx}M.-F. Danca, N. Lung, Appl. Math. Comput. \textbf{223}, 101 (2013).
	\bibitem{mmo} M. Desroches, J. Guckenheimer, B. Krauskopf, C. Kuehn, H.M. Osinga, and M. Wechselberger, SIAM Review \textbf{54}, 211 (2012).

\bibitem{ek}J.-P. Eckmann and D. Ruelle, Rev. Mod. Phys. 57, 617 (1985).

\bibitem{eis} J. Eisenberg, and D. Maszle, J. Theor. Biol. \textbf{176}, 501 (1995).

	\bibitem{el} C. Elton, and M. Nicholson, J Anim Ecol \textbf{11}, 215 (1942).

	\bibitem{foi} C. Foias, and M.S. Jolly, Nonlinearity  \textbf{8}, 295 (1995).

	\bibitem{gom} A.A. Gomes, E. Manica, and M.C. Varriale, Chaos Solitons Frac. \textbf{35}, 432 (2008).

	\bibitem{hai} E. Hairer, S. Norsett, and G. Wanner, \emph{Solving Ordinary Differential Equations I: Nonstiff Problems} (Springer Ser Comput Math, Vol. 1, Springer, 2000).

	\bibitem{hal} J.K. Hale, Bull. Amer. Math. Soc. (N.S.) Mathematical Surveys and Monographs Vol. no. \textbf{25}. Amer Math Soc Providence R. I. (1988).

	\bibitem{par2} G.P. Harmer, and D. Abbott, Stat Sci \textbf{14}, 206 (1999).

	\bibitem{par1} G.P. Harmer, and D. Abbott, Nature \textbf{402}, 864 (1999).

	\bibitem{hast} A. Hastings, and T. Powell, Ecology \textbf{72}, 896 (1991).

\bibitem{hast2} K. McCann and A. Hastings, Proc. R. Soc. London Ser. B: Biol. Sci. \textbf{264}, 1249 (1997).

	\bibitem{hol} C.S. Holling, Mem. Entomol. Soc. Can. \textbf{45}, 1 (1965).

	\bibitem{kleb} A. Klebanoff, and A. Hastings, J. Math. Biol. \textbf{32}, 427 (1994).
	\bibitem{kus} Y. Kuznetsov, O. De Feo, and S. Rinaldi, SIAM J. Appl. Math. \textbf{62}, 462 (200)1

	\bibitem{lot2} A.J. Lotka,  Proc. Natl. Acad. Sci. \textbf{6}, 410 (1920).

	\bibitem{lot1} A.J. Lotka, \emph{Elements of Physical Biology} (Williams and Wilkins Co., 1925).

\bibitem{mai} D.O. Maionchi, S.F. dos Reis, and M.A.M. de Aguiar, Ecol. Model. \textbf{191}291, (2006).
	
\bibitem{mao} Y. Mao, W.K.S. Tang, and M.-F. Danca, Appl. Math. Comput. \textbf{217}, 355 (2010).

	\bibitem{masel} J. Maselko, and H.L. Swinney, Phys Lett A \textbf{119}, 403 (1987).

	\bibitem{may} R.M. May, Ecology \textbf{54}, 638 (1963).

	\bibitem{yod2} K. McCann, and P. Yodzis, Ecology, \textbf{75}, 561 (1994).

	\bibitem{mc} K. McCann, and A. Hastings, Proc. Roy. Soc. Lond. B: Biol. Sci \textbf{264}, 1249 (1997).

	\bibitem{od} E. Odum, \emph{Fundamentals of Ecology} (Saunders Company, 1953).

	\bibitem{og} E. Ott, C. Grebogi, and J.A. Yorke, Phys Rev Lett \textbf{64}, 1196 (1990).

\bibitem{pal} N. Pal, S. Samanta, and J. Chattopadhyay, Chaos, Solitons \& Fractals, \textbf{66}, 58 (2014).

	\bibitem{perk} L. Perko, \emph{Differential Equations and Dynamical Systems }(Springer-Verlag, 1991).

	\bibitem{rin} S. Rinaldi, S. DalBo, and E. DeNittis, J. Math. Biol. \textbf{35}, 158 (1996).

	\bibitem{ros} M.L. Rosenzweig, and R.H. MacArthur, Am. Nat. \textbf{97}, 209 (1963).

	\bibitem{ros2} M.L. Rosenzweig, Science \textbf{171}, 385 (1971).

    \bibitem{rux1} G. Ruxton, Proc. R. Soc. London. Ser. B, \textbf{256}, 189 (1994).

    \bibitem{rux2} G. Ruxton, Ecology \textbf{77}, 317 (1996).

\bibitem{chat4} S. Samanta, T. Chowdhury and J. Chattopadhyay, J. Biol. Phys. \textbf{39}, 469 (2013).

	\bibitem{ver} J.A. Sanders, and F. Verhulst, \emph{Averaging Methods in Nonlinear Dynamical Systems} (Springer-Verlag, 1985).

    \bibitem{sel}E.T. Seton, \emph{The Book of Woodcraft and Indian Lore} (Doubleday, Page \& Company, 1912).
	
    \bibitem{sm} H.S. Smith, \emph{Trans 4th Internatl Congr Ent }\textbf{2}, 191 (1929).

	\bibitem{ston} L. Stone, and D. He, J. Theor. Biol. \textbf{248}, 382 (2007).

	\bibitem{str} S.H. Strogatz, \emph{Nonlinear Dynamics and Chaos: With Applications to Physics, Biology, Chemistry, and Engineering}, (Cambridge University press, 2000).

	\bibitem{xia} X. Tang, Y. He, I.R. Epstein, Q. Wang, S. Wang, and Q. Gao, CHAOS \textbf{24}, 023109 (2014).

\bibitem{vass}D.A. Vasseur, \emph{Biological Chaos and Complex Dynamics, }Oxford Bibliographies in Ecology, (Ed. Gibson D, Oxford University Press, 2012).

	\bibitem{volt} V. Volterra, \emph{Variations and fluctuations of the number of individuals in animal species living together}. (Ed. R.N. Chapman, Animal Ecology, McGraw-Hill, 1931).

	\bibitem{vot} V. Votruba, P, Koubsk\'{y}, D. Kor\v{c}kov\'{a}, and F. Hroch,
 Astronomy \& Astrophysics \textbf{496}, 217 (2009).

\bibitem{xu} C. Xu, and Z. Li, Ecol. Model. \textbf{155}, 71 (2002).
	\bibitem{yod1} P. Yodzis, and S. Innes, Am. Nat., \textbf{139}, 1151 (1992).


\end{thebibliography}
\end{document}